# Lateral Migration and Nonuniform Rotation of Biconcave Particle Suspended in Poiseuille Flow[*]

WEN Bing-Hai(闻炳海)[1,2,3], CHEN Yan-Yan (陈艳燕)[4], ZHANG Ren-Liang(张任良)[1], ZHANG Chao-Ying(张超英)[3,**], FANG Hai-Ping(方海平)[1]

[1] Shanghai Institute of Applied Physics, Chinese Academy of Sciences, Shanghai 201800, China

[2] University of Chinese Academy of Sciences, Beijing 100049, China

[3] College of Computer Science and Information Engineering, Guangxi Normal University, Guilin 541004, China

[4] Physics Department, Zhejiang Normal University, Jinhua 321004, China

A biconcave particle suspended in a Poiseuille flow is investigated by the multiple-relaxation-time lattice Boltzmann method with the Galilean-invariant momentum exchange method. The lateral migration and equilibrium of the particle are similar to the Segré-Silberberg effect in our numerical simulations. Surprisingly, two lateral equilibrium positions are observed corresponding to the releasing positions of the biconcave particle. The upper equilibrium positions significantly decrease with the growth of the Reynolds number, whereas the

---

[*] Supported by the National Science Foundation of China under grant No 10825520 and No 11162002, and the National Basic Research Program of China under grant No 2012CB932400.

[**] Corresponding author.   Email: zhangcy@gxnu.edu.cn

**lower ones are almost insensitive to the Reynolds number. Interestingly, the regular wave accompanied by nonuniform rotation is exhibited in the lateral movement of the biconcave particle. It can be attributed to that the biconcave shape in various postures interacts with the parabolic velocity distribution of the Poiseuille flow. A set of contours illustrate the dynamic flow field when the biconcave particle has successive postures in a rotating period.**

The phenomena of inertia-induced cross-stream migration of suspended particles in Poiseuille flow receive widely interests since the classical investigations,[1] which reported that neutrally buoyant spheres in a pipe flow would migrate away from the wall and reach a certain lateral equilibrium position, namely the Segré-Silberberg effect. After many theoretical, experimental and numerical efforts were made to investigate and analyze the phenomena,[2-7] recently researching interests turn to the biological flows, especially the movement of cells or colloid particles in a tube flow.[8-14] The lateral migration of vesicles was considered as the interplay between nonlinear character of a Poiseuille flow and vesicle deformation.[9] Soon, theoretical analysis ascribed the cross-stream shift to the ratio of the inner over the outer fluid viscosities.[10] The more recent studies observed that vesicles presented two motion patterns (oscillation and vacillating breathing)[13] or a phase diagram of shapes (bullet, croissant and parachute).[14] Red blood cell of human being could be the most famous vesicle in biological flows. Without a cell nucleus, it generally exhibits a biconcave shape in rest state and some viscoelastic deformations in blood flow.[15-17] However, sizeable disparities can still be noticed among numerical simulations and experimental observations due to its tiny size and complex characters.[13-20] It is therefore meaningful to obtain some credible benchmarks by simple models in order to understand the essential behaviors of particles with various geometries. For example, the investigations of ellipse in sedimentations,[21] shear flow[22] and Poiseuille flow[5] can make active promotions to the researches of blood circulation of birds, whose red blood cell is oval or elliptical in shape with a nucleus inside.[23]

Here, we apply a simple model to investigate a biconcave particle suspended in a two-dimensional Poiseuille flow by using the multiple-relaxation-time lattice Boltzmann method (LBM)[24-27] with the Galilean-invariant momentum exchange method.[28] Surprisingly, different from the general observations, two lateral equilibrium positions are found for the biconcave particle and Reynolds numbers play different roles in the lower and the upper equilibriums. Orientations, regular wave and nonuniform rotation of the particle are presented in detailed. Furthermore, the dynamic flow fields around the particle are illustrated vividly by the contours of the vertical velocities when the particle has successive postures.

The lattice Boltzmann equation with multiple relaxation times can be written as[26, 27]

$$f_i(\boldsymbol{x}+\boldsymbol{e}_i\delta t, t+\delta t) - f_i(\boldsymbol{x},t) = -\mathrm{M}^{-1}\cdot\mathrm{S}\cdot\left[\mathbf{m}-\mathbf{m}^{(eq)}\right], \tag{1}$$

where $f_i(\boldsymbol{x},t)$ is the particle distribution function at lattice site $\boldsymbol{x}$ and time $t$, moving along the direction defined by the discrete speeds $\boldsymbol{e}_i$, and $\delta t$ is the time step. $\mathbf{m}$ and $\mathbf{m}^{(eq)}$ represent the velocity moments of the distribution functions and their equilibria, respectively. For the model with two dimensions and nine discrete velocities, $i$ is an integer $0 \leq i \leq 8$ and the velocity moments are $\mathbf{m}=(\rho, e, \varepsilon, j_x, q_x, j_y, q_y, p_{xx}, p_{xy})^\mathrm{T}$. The conserved moments are the density $\rho$ and the flow momentum $\boldsymbol{j}=(j_x, j_y)=\rho\boldsymbol{u}$, $\boldsymbol{u}$ is the local velocity vector. The equilibria

of nonconserved moments depend only on the conserved moments:

$$e^{(eq)} = -2\rho + \frac{3}{\rho}(j_x^2 + j_y^2), \quad \varepsilon^{(eq)} = \rho - \frac{3}{\rho}(j_x^2 + j_y^2), \quad q_x^{(eq)} = -j_x, \quad q_y^{(eq)} = j_y,$$

$p_{xx}^{(eq)} = \frac{1}{\rho}(j_x^2 - j_y^2), \ p_{xy}^{(eq)} = \frac{1}{\rho}(j_x j_y)$. M is a linear transformation matrix mapping between moment space and discrete velocity space, $\mathbf{m} = \mathbf{M} \cdot \mathbf{f}$ and $\mathbf{f} = \mathbf{M}^{-1} \cdot \mathbf{m}$. S is a diagonal matrix of nonnegative relaxation rates and is given by $\mathbf{S} = \text{diag}(0, s_e, s_\varepsilon, 0, s_q, 0, s_q, s_v, s_v)$. Then the shear viscosity is $v = \frac{1}{3}\left(\frac{1}{s_v} - \frac{1}{2}\right)$.

The hydrodynamic force can be evaluated simply and efficiently by the momentum exchange method in the lattice Boltzmann method. In the conventional schemes, the hydrodynamic force on a single fluid-solid link is computed by[29-32]

$$\mathbf{F}(\mathbf{x}_s) = \mathbf{e}_i f_i(\mathbf{x}_f, t) - \mathbf{e}_{\bar{i}} f_{\bar{i}}(\mathbf{x}_b, t), \tag{2}$$

where $\mathbf{x}_f$ and $\mathbf{x}_b$ are a fluid node and a boundary node on a fluid-solid link which intersects the boundary at $\mathbf{x}_s$. However, the schemes result in sizeable errors when a simulation involves moving boundaries.[32, 33] Recently, Wen *et al.* proposed a Galilean-invariant momentum exchange method by introducing the relative velocity into the interfacial momentum transfer to compute the hydrodynamic force[28]

$$\mathbf{F}(\mathbf{x}_s) = (\mathbf{e}_i - \mathbf{v}) f_i(\mathbf{x}_f, t) - (\mathbf{e}_{\bar{i}} - \mathbf{v}) f_{\bar{i}}(\mathbf{x}_b, t). \tag{3}$$

It is demonstrated to greatly enhance the hydrodynamic accuracy and robustness of moving boundaries in dynamic fluid. Especially, the algorithm meets full Galilean invariance and is independent of boundary geometries. The total hydrodynamic force and torque acting on the solid particle are evaluated by:

$$\mathbf{F} = \sum \mathbf{F}(\mathbf{x}_s) \quad \text{and} \quad \mathbf{T} = \sum (\mathbf{x}_s - \mathbf{R}) \times \mathbf{F}(\mathbf{x}_s), \tag{4}$$

where $R$ is the mass center of the solid particle, and the summation runs over all the fluid-solid links.

Figure 1 illustrates a schematic diagram of a biconcave particle suspended in a Poiseuille flow. Both densities of the fluid and the particle are 1 g/cm$^3$. The width of the channel is 0.1 cm and the kinematic viscosity is $\upsilon$=0.01 cm$^2$/s. The Reynolds number is defined to characterize the flow domain by $Re = HU/\upsilon$, where $U$ is the mean velocity of the Poiseuille flow without particle and $H$ is the width of the channel. The biconcave shape of a red blood is described by the following equation which was given by Fung *et al*. in 1980s[34]

$$y = \frac{1}{2}\left[1-(\frac{x}{R})^2\right]^{1/2}\left[C_0 + C_1(\frac{x}{R})^2 + C_2(\frac{x}{R})^4\right], \tag{5}$$

where $C_0 = 0.81$, $C_1 = 7.83$, $C_2 = -4.39$ and $R = 2.91$. In the present simulations, the relaxation rates are set to $s_e = 1.64$, $s_\varepsilon = 1.54$, $s_q = 1.9$ and $s_\nu = 1/0.6$. The width of the computational domain is $H$=100 lattice units while the length is 20 times the width. The particle radius is $R$=15 lattice units. The second-order interpolation boundary condition[35] is adopted for the channel wall and the particle boundary. The pressure boundary condition[36] is applied at both the inlet and outlet of the channel. With the parallel optimization of Intel OpenMP, a simulation which contains 60 seconds of particle movement performs 18 million time steps and takes about 60 hours on a HP Z600 with 12 cores inside.

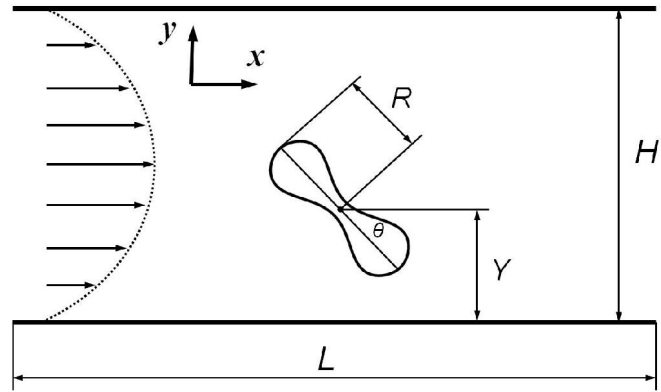

*Fig. 1.* *Schematic diagram of a biconcave particle suspended in a Poiseuille flow.*

The biconcave particles are released at 0.02 and 0.04 cm in the low half of the flow field and at the center of the channel in the horizontal direction. The lattices will be redrawn when the particle moves more than two lattice lengths in the horizontal direction in order to keep the particle in the middle of the channel all the time. Fig. 2 draws the trajectories of the particle at the Reynolds numbers 3 and 12, respectively. The lateral migration and equilibrium with periodic wave are clearly observed. We also present the trajectories of the classic Segré-Silberberg effect with the circular particle of radius 15 as comparisons. As shown in Fig. 2, it is usually agreed that a single lateral equilibrium position is located between the wall and the centerline of the channel for the classic Segré-Silberberg effect.[2, 7, 9, 13] Surprisingly, two lateral equilibrium positions are exhibited for biconcave particles and they locate at both sides of the equilibrium position of the circular particle. The smaller the Reynolds number is, the more slowly the particle reaches the equilibrium and the farther the two equilibrium positions separate.

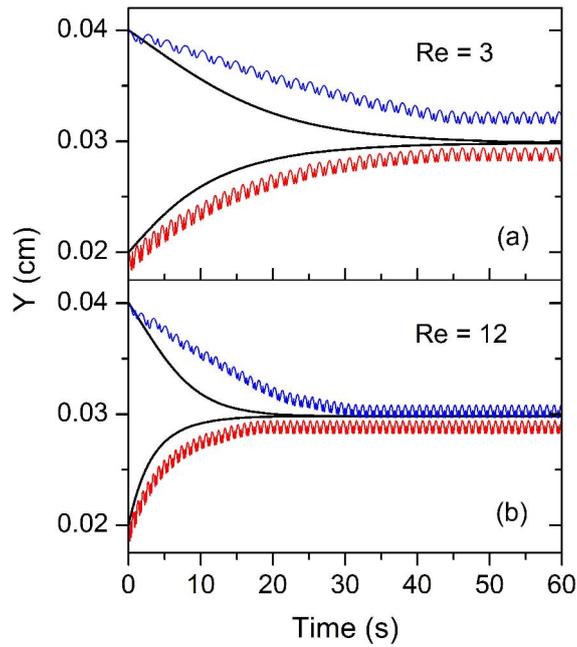

***Fig. 2.*** *The migrating trajectories of a biconcave particle in a Poiseuille flow with two Reynolds numbers (a) Re=3 and (b) Re=12. The particles in the red and the blue trajectories are released at 0.02 and 0.04cm away from the low wall, respectively. The black trajectories represent the classic Segré-Silberberg effect in which the particle is a circle.*

These trends can be seen more clearly in Fig. 3(a), in which a set of Reynolds number are simulated. The upper equilibrium positions become low significantly with the increase of the Reynolds number, whereas the lower ones are almost insensitive to the Reynolds number. Please note that the horizontal axis in Fig. 3 uses a binary logarithmic coordinate.

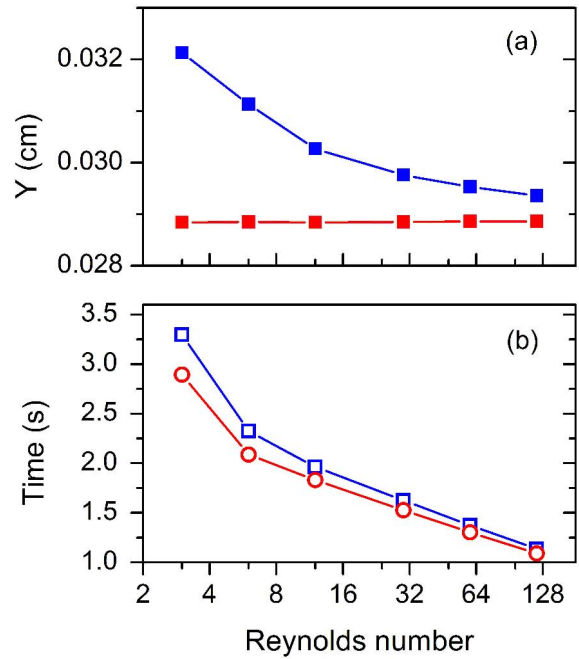

*Fig. 3. The impacts of Reynolds numbers on (a) the final equilibrium positions and (b) the rotating period of the particle. The particles in the red and blue lines are released at 0.02 and 0.04cm away from the low wall, respectively.*

Although the biconcave particle in a Poiseuille flow behaves like the Segré-Silberberg effect, it includes additional regular wave and nonuniform rotation due to the noncircular geometry. Explicitly, we define a rotating period as a time interval in which a biconcave particle rotates a round. As shown in Fig. 3(b), the rotating periods are monotonic decreasing with the increase of the Reynolds numbers. Notably, the rotating periods of the upper equilibrium positions are always longer than the lower ones and the gaps between them show a continuous narrowing. These occur as a result of the parabolic velocity distribution of a Poiseuille flow. The speeds of flow at the upper positions are higher than those at the lower ones, and the velocity differences reduce continuously since they approach each other along with the growth of the Reynolds numbers.

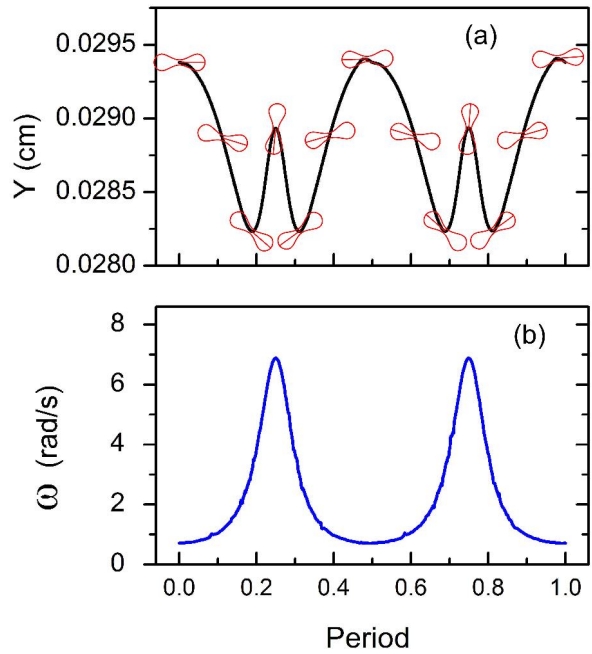

*Fig. 4.* *(a) The trajectory and orientation and (b) the angle velocity of the biconcave particle in a single rotating period.*

The trajectory and angle velocity of the biconcave particle in a single period in equilibrium state are illustrated in Fig. 4, together with the orientations of the particle in various typical positions. The Reynolds number is 3 and the particle starts from 0.02 cm away from the low wall and rotates clockwise. The migrating trajectory reaches the ridges when the particle angle is about $0\pi$ and $1\pi$, and two steeper sub-ridges at about $0.5\pi$ and $1.5\pi$. Noticeably, the ridges match the minimum angle velocities whereas the sub-ridges match the maximum ones. The peaks of the trajectory have a little lag relative to the angle of the biconcave particle. Four similar troughs lie about $0.25\pi$, $0.75\pi$, $1.25\pi$ and $1.75\pi$. Their corresponding angle velocities are moderate but change fast.

The mechanism of lateral migration in the Segré-Silberberg effect is usually explained by the inertia effect[3, 7]. The additional wave and nonuniform rotation in our simulations are due to the interaction of the biconcave shape and the parabolic velocity distribution. The fluid velocity at the upper part of the particle is always faster than that at the lower part and the difference drives the particle to rotate incessantly. The changes of the posture would impact the hydrodynamic forces exerted by the fluid flow. It can be indicated in Fig. 4 that the biconcave particle is exerted a dropping force in the about angle range of 0 ~ 0.25π and a lifting force in the about angle range of 0.75 ~ 1π. The flow field around the biconcave particle is carefully investigated by the following contour diagrams.

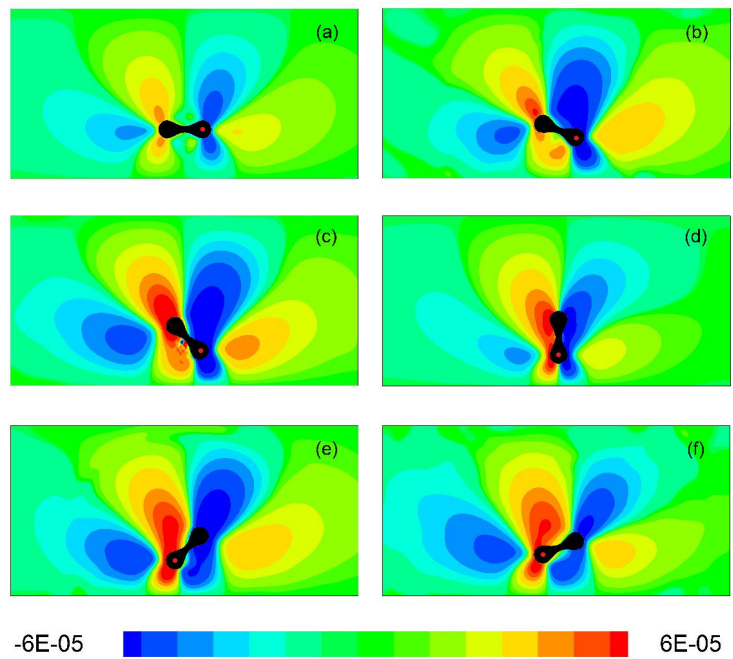

*Fig. 5. The contours of vertical velocities of the fluid around the biconcave particle. The subfigures (a) ~ (f) are corresponding to the first six positions in Fig. 4(a).*

Since a Poiseuille flow is a well-defined laminar flow, the vertical velocity of the present flow field originates totally from the wave and rotation of the biconcave particle and is far more sensitive to the particle motion than the horizontal velocity. Therefore we apply the contour of the vertical velocity to characterize the flow field. The subfigures (a) ~ (f) are the counterparts of the first six locations in Fig. 4(a). The subfigure (a) is at the highest position with a horizontal posture and the smallest angle velocity, the flow is mild and bilaterally balanced. The subfigures (c) and (e) are at the troughs with sloping postures, the flows are strong. The subfigures (b) and (f) draw the rising and falling stages, and thus the upward and downward flows dominate the flow field, respectively. The subfigures illustrate the successive changes and depict the dynamic flow field vividly.

In this work, we perform a series of numerical simulations of a biconcave particle migrating laterally in a Poiseuille flow by the lattice Boltzmann method with multiple relaxation times. The hydrodynamic force is evaluated by the Galilean-invariant momentum exchange method. Because of the biconcave shape, the particle finally reaches the different equilibrium positions corresponding to its releasing points. This makes a remarkable distinction to the classic the Segré-Silberberg effect, in which a single equilibrium position is observed for a circle or sphere. Extending the simulations to a range of Reynolds numbers, we find that the upper equilibrium positions significantly decrease with the growth of the Reynolds number but the lower ones are almost insensitive to the Reynolds number. Inside a single rotating period, the biconcave particle moves with regular wave and the nonuniform angle velocity. The around flow field is highly dynamic when the particle takes different postures. These can be attributed to the interaction of the parabolic velocity distribution and the biconcave shape. The observed phenomena indicate the complex motions of red blood cells migrating in a blood vessel, in which blood flow involves non-Newtonian characters and periodic pulse. Accompanied by deformation into a slipper-shape and tank tread-like motion,[17-19] we conjecture that a red blood cell would rotate and wave in a Poiseuille flow together with more than one equilibrium state. These can be numerically presented by accurate force evaluation, and the lattice Boltzmann method with the Galilean-invariant momentum exchange method can be a promising candidate. The investigation will be expanded to combine with the viscoelastic membrane[15] in order to enrich the understanding of the behaviors of red blood cell and other vesicles in dynamic fluid.

The authors thank Shanghai Supercomputer Center of China for the support of computation.